\documentclass[aps,prl,nofootinbib,twocolumn,superscriptaddress,preprintnumbers,balancelastpage,longbibliography,nobibnotes]{revtex4-2}

\usepackage{amsmath,amssymb} 
\usepackage{graphicx}
\usepackage{hyperref}
\usepackage{enumitem}
\usepackage{xcolor}
\usepackage{tikz-cd}

\def\O{\mathcal{O}}
\def\L{\mathcal{L}}
\def\Tr{\text{Tr}}
\def\id{\mathbb{I}}
\def\Z{\mathbb{Z}}
\def\hc{\text{h.c.}}
\def\U32{$U(3) \times U(2)$}

\definecolor{aqua}{rgb}{0.4, 0.6, 0.7}

\newcommand{\AddrIFIC}{%
  Instituto de F\'{i}sica Corpuscular, CSIC-Universitat de Val\`{e}ncia, 46980 Paterna, Spain}

\newcommand{\AddrFISTEO}{%
  Departament de F\'{i}sica Te\`{o}rica, Universitat de Val\`{e}ncia, 46100 Burjassot, Spain}
\begin{document}

%-------------------
% Title and authors
%-------------------
\title{Beyond $\boldsymbol{SU(N)}$:\\
$\boldsymbol{U(3) \times U(2)}$ as the underlying symmetry of the strong and electroweak interactions}

\author{Antonio Herrero-Brocal}
\email{antonio.herrero@ific.uv.es}
\affiliation{\AddrIFIC, Valencia, Spain}

\author{Javier Perez-Soler}
\email{javier.perez.soler@ific.uv.es}
\affiliation{\AddrIFIC, Valencia, Spain}

\author{Avelino Vicente}
\email{avelino.vicente@ific.uv.es}
\affiliation{\AddrIFIC, Valencia, Spain}
\affiliation{\AddrFISTEO, Valencia, Spain}

%-------------------
% Abstract
%-------------------
\begin{abstract}
The gauge principle is a cornerstone of particle-physics model building. Nevertheless, many constructions leave certain global $U(1)$ redundancies ungauged. In this work, we take the gauge principle to its logical extreme by promoting all $SU(N)$ symmetries to $U(N)$. We focus on a model based on local $U(3)\times U(2)$ invariance. This framework accounts for several otherwise \textit{ad hoc} features of the Standard Model, including charge quantization and the observed hypercharge assignments, which emerge here as unique predictions. Furthermore, the internal consistency of the model requires the introduction of right-handed neutrinos and implies the presence of an additional $U(1)$ factor that can be identified with $B-L$, thereby naturally yielding non-zero neutrino masses. In light of these findings, we hypothesize that $U(3)\times U(2)$ constitutes the underlying symmetry of the strong and electroweak interactions. More importantly, our approach opens up novel avenues for model building, driven by this extended interpretation of the gauge principle.
\end{abstract}

\maketitle

\noindent \textbf{Introduction.} Symmetry lies at the heart of the Standard Model (SM), a Yang–Mills theory with local gauge invariance~\cite{Glashow:1961tr,Weinberg:1967tq,SalamBook}. Its architects sought the minimal framework capable of explaining all known phenomena in particle physics at the time. In doing so, a fundamental symmetry structure was identified, virtually always associated with the group $SU(3)_c \times SU(2)_L \times U(1)_Y$. Remarkably, the model built upon this symmetry successfully described both the strong and electroweak interactions and predicted the existence of new particles, the gluons, and the $W$, $Z$ and Higgs bosons, all of which were eventually discovered in particle colliders.

While symmetries impose many restrictions on physical systems, they also leave room for free choices. This is also true for the SM. In fact, some of its fundamental ingredients are completely \textit{ad hoc}. For instance, the hypercharge assignments have no rationale within the SM itself, but are simply assumed. As a consequence of these assignments, one finds that the magnitudes of the electric charges of the electron and the proton happen to be exactly the same, even though this equality does not follow from any deeper principle. Moreover, some minimal representations are absent from the SM (and unobserved in Nature), despite posing no difficulty from a group–theoretical point of view. This is the case, for example, of an $SU(2)_L$ doublet with vanishing hypercharge. Taken together, these features make the SM a remarkably accurate quantum field theory, but one built on the basis of empirical inputs rather than on an underlying organizing principle. 

Gauge symmetries are redundancies in our description of physical systems. A well-known example is quantum electrodynamics: the $U(1)$ symmetry reflects the redundancy of degrees of freedom in the vector potential. The same holds in non-Abelian theories, where local $SU(N)$ transformations encode the redundancy. However, when constructing an $SU(N)$ theory, one typically encounters an additional global $U(1)$ symmetry arising from the fact that fermions enter the Lagrangian through bilinear combinations. This naturally raises the question: why should we gauge only the $SU(N)$ group while leaving the associated phase as a global symmetry?

In this work \textbf{we fully embrace the gauge principle and promote all $\boldsymbol{SU(N)}$ groups to $\boldsymbol{U(N)}$}. More precisely, we consider \U32 as the underlying local symmetry behind the SM dynamics. We will show that this choice not only reproduces the SM, but also explains many of its ad hoc features. The group structure readily implies:
\begin{enumerate}[label=(\roman*)]
\item Charge is quantized.
\item Some representations are forbidden. In particular, exotic ones such as the $SU(2)_L$ doublet with vanishing hypercharge.
\end{enumerate}
In addition, when combined with anomaly cancellation, a condition that is required for consistency at the quantum level, one automatically finds:
\begin{enumerate}[label=(\roman*)]
\item[(iii)] The SM hypercharge symmetry is predicted. The only solution consistent with anomaly cancellation turns out to be the one leading to the SM Lagrangian.
\end{enumerate}
These results would be equally valid if the SM was based on the gauge groups $SU(3) \times U(2)$, $U(3) \times SU(2)$ or $S\left( U(3) \times U(2) \right)$. The fact that the SM Lagrangian is compatible with different gauge groups has already been discussed in~\cite{ORaifeartaigh:1986agb,Hucks:1990nw,Tong:2017oea}. However, such constructions generally fail once the theory is extended beyond the SM. For instance, upon introducing a right-handed neutrino, the hypercharge assignment is no longer determined. Our idea goes beyond identifying gauge groups that are merely consistent with the SM. Instead, we hypothesize that any $SU(N)$ must be accompanied by a global phase redundancy, thereby promoting $SU(N)$ to $U(N)$. This leads to $U(3)\times U(2)$ and makes the following predictions:
%\ahb{
%These claims coming from the group structure would remain equally valid if the SM were described by the gauge groups $SU(3)_c \times U(2)$, $U(3) \times SU(2)_L$ or $S\left( U(3) \times U(2) \right)$. The fact that different global groups can give rise to the same SM~\footnote{This observation illustrates that the SM is a model based on the Lie algebra $\mathfrak{su}(3) \oplus \mathfrak{su}(2) \oplus \mathfrak{u}(1)$ rather than the group $SU(3)_c \times SU(2)_L \times U(1)_Y$.} has already been discussed in~\cite{ORaifeartaigh:1986agb, Hucks:1990nw, Tong:2017oea}. These groups are constructed to reproduce the SM and indeed they do. However, such constructions generally fail once the theory is extended beyond the SM (BSM). For instance, upon introducing a right-handed neutrino, the hypercharge assignment is no longer determined. Our idea in this work goes beyond identifying gauge groups that are merely consistent with the SM. We hypothesize that any $SU(N)$ must be accompanied with a global phase redundancy, thereby promoting $SU(N)$ to $U(N)$. Therefore, our approach impose $U(3)\times U(2)$ leading to BSM predictions:
%}

%The fact that these features are fixed by our construction strongly suggests that $\boldsymbol{U(3) \times U(2)}$ \textbf{should be regarded as the genuine underlying symmetry of the strong and electroweak interactions}. In addition, our setup makes the following predictions:
%
\begin{enumerate}[label=(\roman*)]
\item[(iv)] $B-L$ is predicted as a gauge symmetry. Therefore, this symmetry is no longer global and accidental, as in the SM, but local and required for consistency.
\item[(v)] Right-handed neutrinos must exist.
\end{enumerate}

Based on these findings, we propose $\boldsymbol{U(3) \times U(2)}$ \textbf{as the genuine underlying symmetry of the strong and electroweak interactions}. \\

\noindent \textbf{Preliminaries.} Before introducing our \U32 model, we highlight a few general results about $U(N)$ groups that will play a crucial role in what follows. An extended discussion and the proofs of the claims made in this Section are provided in the End Matter.

Every Lie group has an associated Lie algebra, essentially specified by its generators and their commutation relations. In particle physics, we usually care about the algebra more than about the group itself, since it is the algebra that determines the form of the interaction terms in the Lagrangian. It is therefore tempting to assume that two Lie groups with the same algebra are equivalent, as they would appear to lead to the same dynamics. However, Lie groups with isomorphic Lie algebras need not be isomorphic. This distinction has been widely overlooked in the literature, even though it is made in some introductory texts~\cite{ORaifeartaigh:1986agb,Zee:2016fuk,Reece:2023czb} and has important consequences for model building.

Let us now consider the case of $U(N)$ and $SU(N) \times U(1)$. Although these two groups have the same generators and share the same algebra, they are different. The true group isomorphism is given by
\begin{equation} \label{eq:iso}
    U(N) \cong \frac{SU(N) \times U(1)}{\mathbb{Z}_N} \; .
\end{equation}
This implies that $SU(N) \times U(1)$ is an $N$-fold cover of $U(N)$, in much the same way $SU(2)$ is a double cover of $SO(3)$. Each element of $U(N)$ corresponds to $N$ elements of $SU(N) \times U(1)$. 

The global structure of a group determines its allowed representations. In particular, because of the isomorphism in Eq.~\eqref{eq:iso}, any representation of $U(N)$ is automatically a representation of $SU(N) \times U(1)$. The converse, however, does not hold: there exist representations of $SU(N) \times U(1)$ that are not representations of $U(N)$. Let \textbf{R}$_q$ denote a representation of $SU(N) \times U(1)$, with \textbf{R} the representation under $SU(N)$ and $q$ the $U(1)$ charge. Then, \textbf{R}$_q$ is also a representation of $U(N)$ only for certain values of $q$. For $N=2$ and $N=3$, the two cases relevant to our work, the restrictions are given by
\begin{align}
    q_2 &= 2 \, j +2 \, n_2 \, , \label{eq:rule2} \\    
    q_3 &= p-q +3 \, n_3 \, , \label{eq:rule3}
\end{align}
where $q_2$ and $q_3$ are the charges under the $U(1)$ factor, the $SU(2)$ and $SU(3)$ representations have Dynkin labels $2 \, j$ (with $j$ the usual isospin of $SU(2)$) and $(p, q)$, respectively, and $n_2, n_3 \in \mathbb{Z}$.

Eqs.~\eqref{eq:rule2} and \eqref{eq:rule3} are central to our work. They constrain the allowed $U(1)$ charges of the $U(N)$ representations to be included in any particle physics model. For this reason, they will be referred to in what follows as \textbf{the charge constraints}. Let us consider some examples. For the fundamental representation of $SU(2)$, the doublet (with $j=1/2$), Eq.~\eqref{eq:rule2} implies that only odd values of $q_2$ are consistent with $U(2)$. This allows \textbf{2}$_1$ but forbids \textbf{2}$_0$. In the case of the fundamental representation of $SU(3)$, the triplet (with $(p,q)=(1,0)$), Eq.~\eqref{eq:rule3} implies that $q_3$ cannot be a multiple of $3$. Similarly, the singlet and fundamental representations of $SU(N)$ cannot carry the same charge under the associated $U(1)$.

Another consequence of $U(N)$ and $SU(N) \times U(1)$ sharing the same algebra is that they have the same covariant derivative. In particular, even though $U(N)$ is a single gauge group, its most general covariant derivative involves two independent gauge couplings: one associated with the $SU(N)$ generators and another with the $U(1)$ generator. More generally, the transformation properties of the covariant derivative are unaffected when two generators have different gauge couplings, provided they commute. This is exactly the case here, since the $U(1)$ generator is proportional to the identity and thus commutes with all other generators. \\

\noindent \textbf{The model.} Our model replaces the SM gauge symmetry group by $U(3)\times U(2)$:
\begin{equation}
    U(3)\times U(2) \cong \left(\frac{SU(3)_c \times U(1)_3}{\mathbb{Z}_3}\right) \times \left(\frac{SU(2)_L \times U(1)_2}{\mathbb{Z}_2} \right) \, .
\end{equation}
The charges under $U(1)_3$ and $U(1)_2$ will be denoted by $q_3$ and $q_2$, respectively, and are constrained to verify Eqs.~\eqref{eq:rule2} and \eqref{eq:rule3}. Therefore, \textbf{charge is quantized and not all the representations are allowed}. This is the first major result of our model.

Imposing anomaly cancellation to our symmetry group (before charge quantization) is equivalent to do it for $SU(3)\times SU(2) \times U(1) \times U(1)$. For a non-Abelian gauge group, this implies
%
%\begin{align}
% \sum_{i=1}^{N_L} \Tr\left(T^a \left\{T^b \, , T^c \right\} \right) \left(\mathbf{R}_{L_i}\right) -  \sum_{i=1}^{N_R} \Tr\left(T^a \left\{T^b \, , T^c \right\} 
% \right)\left(\mathbf{R}_{R_i}\right)= 0 \; ,
%\end{align}
\begin{align}
 \sum_{i=1}^{N_L} \mathcal{A}^{abc} \left(\mathbf{R}_{L_i}\right) -  \sum_{i=1}^{N_R} \mathcal{A}^{abc} \left(\mathbf{R}_{R_i}\right)= 0 \; ,
\end{align}
where $\mathcal{A}^{abc} = \Tr\left(T^a \left\{T^b \, , T^c \right\} \right)$ is the usual symmetrized trace over the generators $T$ of the groups and the sums extend over the left- and right-handed fermions with non-trivial representations \textbf{R}. In our case, the cubic and gravity anomalies lead to
\begin{align}
    U(N)^3 \to 
   \left\{ \begin{array}{cl}
SU(N)^3 \\
SU(N)^2 \, U(1) \\
U(1)^3
\end{array} \right. , \, G^2 \, U(N) \to G^2 \, U(1) \, .
\end{align}
We assume that all fermions in the model belong to either the fundamental or the singlet representation under $SU(N)$. Under this assumption, together with the charge constraints, the equation arising from the $SU(3)^2 \, U(1)_3$ anomaly leads to $N_{3L} = N_{3R} + 3 k$, with $k \in \Z$ and $N_{3L}$ and $N_{3R}$ the number of left and right-handed $SU(3)$ fundamentals, respectively. In the same way, the $SU(2)^2 \, U(1)_2$ anomaly leads to $N_{2L}= 2 k$, with $k\in \Z$, matching the Witten anomaly. Moreover, the minimal scheme requires two singlets under both $SU(3)$ and $SU(2)$. Therefore, the minimal fermion content coincides with that of one SM generation with the addition of a new singlet. The complete fermion content and charge assignments are summarized in Table~\ref{tab:Fcontent}.
\begin{table}[t!]
\centering
\begin{tabular}{|c|c|c|c|c|}
\hline
Fields & $SU(3)_c$ & $SU(2)_L$ & $U(1)_3$ & $U(1)_2$  \\ \hline
$Q_L $  & \textbf{3} & \textbf{2} & $q_3$ & $q_2$ \\ \hline
$u_R$   & \textbf{3} & \textbf{1} & $u_3$ & $u_2$  \\ \hline
$d_R$   & \textbf{3} & \textbf{1} & $2 \, q_3 - u_3$ &  $2 \, q_2 - u_2$ \\ \hline
$L_L $  & \textbf{1} & \textbf{2} & $-3 \, q_3$ &  $-3 \, q_2$\\ \hline
$e_R$   & \textbf{1} & \textbf{1} & $-4 \, q_3 +u_3$ & $-4 \, q_2 +u_2$  \\ \hline
$\nu_R$   & \textbf{1} & \textbf{1} & $-2 \, q_3 -u_3$ & $-2 \, q_2 - u_2$  \\ \hline
\end{tabular}
\caption{Minimal fermionic content of $U(3) \times U(2)$. The charges under the $U(1)$ factors are given in the last two columns in terms of the charges of $Q_L$ ($q_i$) and $u_R$ ($u_i$), as required by anomaly cancellation. In addition, we note that $q_i$ and $u_i$ must verify the charge constraints in Eqs.~\eqref{eq:rule2} and \eqref{eq:rule3}. In this case, the charge constraints for the rest of the fermions in the model are automatically satisfied.
\label{tab:Fcontent}
}
\end{table}
These charges, together with the charge constraints lead to some important consequences:
\begin{itemize}
     \item \textbf{Charge is quantized}. The solution to the anomaly cancellation conditions is obtained independently of charge quantization. This implies that the same relations among the charges would be found if one were to add an extra singlet to the SM. In that case, however, the ratio $u_i/q_i$ is not necessarily in $\mathbb{Q}$ and hypercharge is not quantized. However, within our framework all charges are quantized by the symmetry, and the observed rational relations among the charges arise automatically.
    \item \textbf{The observed hypercharge is predicted uniquely in this framework}. In the SM, anomaly cancellation admits more than one solution for the charges under the $U(1)$ factor, since both the trivial ($Y_i =0 \, , \forall \, i \in \text{SM}$) and vector-like ($Y_Q = Y_L=Y_e= 0$ and $Y_u= -Y_d$) assignments are always anomaly-free possibilities. In our scheme these two solutions are ruled out since they are not compatible with the charge constraints. Moreover, there always exists a $U(1)$ combination that reproduces the observed hypercharge gauge symmetry
     \begin{equation}
        Y = -\frac{\left( 2 \, q_2 + u_2 \right) Q_3 - \left( 2 \, q_3 + u_3 \right) Q_2}{6 \left( q_2 \, u_3 - q_3 \, u_2 \right)}  \; ,
     \end{equation}
    with $Q_i$ the generator of the $U(1)_i$ symmetry group. We note that the choice $q_2 \, u_3 - q_3 \, u_2 = 0$ would lead to identical $U(1)_3$ and $U(1)_2$ factors, up to normalization, rendering the $U(3)\times U(2)$ construction redundant. Thus, this generator always reproduces the SM hypercharge and predicts the new fermion to be a singlet, i.e., $Y = 0$. In other words, \textbf{the model predicts the existence of the right-handed neutrino}.
    \item \textbf{The observed $\boldsymbol{B-L}$ is predicted uniquely as a gauge symmetry}. Again, one can always find a $U(1)$ combination that reproduces the observed $B-L$ charge assignment.  Interestingly, there exists a particular assignment in which $U(1)_3$ and $U(1)_2$ corresponds to $B-L$ and $Y$, respectively, leading to the classical scenario in which the new gauge boson couples according to the $B-L$ charges. It is also important to emphasize that the converse is not valid: there is no consistent assignment that maps $U(1)_2$ to $B-L$. 
    \item \textbf{A new heavy $Z'$ gauge boson is predicted}. Since all SM fields are charged under the additional $U(1)$, its associated gauge boson must be sufficiently heavy to avoid phenomenological problems. This new $Z'$ boson couples to the hypercharge-orthogonal combination, which is not necessarily associated with $B-L$. Consequently, the new interaction need not have the same strength for all quarks or all leptons. Nevertheless, the relations listed in Tab.~\ref{tab:Fcontent} hold for any linear combination of the individual $U(1)$ factors. For example, the interaction strength of the new gauge boson with $L$ must be three times that with $Q$, while its interaction with $d_R$ is determined by the charge relation $d_i = 2 q_i - u_i$.
    %
    %This new $Z'$ boson couples to the hypercharge orthogonal combination, but the relation between the charges remains invariant. Let us emphasize this point. Since our $U(1)$ charges are not imposed by hand, the new interaction is not necessarily associated with $U(1)_{B-L}$. For instance, the interaction with the quarks could differ for each one while still preserving the overall charge assignment, $d =2 q - u $.
 %   
\end{itemize}
\begin{table}[t!]
\centering
\begin{tabular}{|c|c|c|c|c|}
\hline
Fields & $SU(3)_c$ & $SU(2)_L$ & $U(1)_3$ & $U(1)_2$  \\ \hline
$H $  & \textbf{1} & \textbf{2} & $h_3$ & $h_2$ \\ \hline
$\Phi$   & \textbf{1} & \textbf{1} & $\phi_3$ & $\phi_2$  \\ \hline
\end{tabular}
\caption{Minimal scalar content of $U(3) \times U(2) $. Due to theoretical considerations, the $U(1)_i$ charges must satisfy Eqs.~\eqref{eq:rule2}  and~\eqref{eq:rule3}. On the other hand, we impose $\left(2 \, q_3 + u_3 \right) \,\phi_2=\left(2 \, q_2 + u_2 \right) \,\phi_3$ and $h_i = u_i -q_i$ in order to reproduce the SM, as explained in the text.
\label{tab:Scontent}}
\end{table}
We have derived the minimal fermion content just from theoretical considerations related to the internal consistency of the theory. The scalar sector, however, must be chosen on phenomenological arguments rather than theoretical ones, since the addition of scalar fields does not affect the consistency of the model. First, at the electroweak scale we must recover the SM. Hence, we require a singlet scalar, $\Phi$, that breaks spontaneously the additional $U(1)$ gauge symmetry. This singlet must be uncharged under $Y$, which translates into the condition $\left(2 \, q_3 + u_3 \right) \,\phi_2=\left(2 \, q_2 + u_2 \right) \,\phi_3$, with $\phi_i$ the $\Phi$ charge under $U(1)_i$. Moreover, a Higgs-like field is needed to break the electroweak symmetry. Note that any scalar that compensates the $U(1)$ charges of, for example, the operator $\bar{Q} u_R$, also does so for $\bar{Q} d_R$, $\bar{L} e_R$ and $\bar{L} \nu_R$. In other words, the SM Yukawa sector is recovered by imposing $h_i = u_i -q_i$,
\begin{align}
    \L_Y = y_d \, \bar{Q} H d_R + y_u \, \bar{Q} \tilde{H} u_R + y_e \, \bar{L} H e_R + y_\nu \, \bar{L} \tilde{H} \nu_R + \hc\, ,
\end{align}
and this is automatically consistent with the charge constraints. Thus, the minimal scalar sector is given in Tab.~\ref{tab:Scontent}.
%Apart from the fermionic sector, the minimal setup requires a singlet scalar, $\Phi$, that spontaneously breaks the additional $U(1)$ gauge symmetry in order to recover the SM at low energies. This singlet must be uncharged under $Y$, which translates into the condition $\left(2 \, q_3 + u_3 \right) \,\phi_2=\left(2 \, q_2 + u_2 \right) \,\phi_3$, with $\phi_i$ the $\Phi$ charge under $U(1)_i$.  This equation implies that all singlet fields charged only under the broken $U(1)$ must carry integer multiples of a minimal charge \av{(is this observation relevant?)}. Moreover, a Higgs-like field is needed to break the electroweak symmetry. Thus, the minimal scalar sector is given in Tab.~\ref{tab:Scontent}.
%
Symmetry breaking proceeds as
\begin{displaymath}
\begin{tikzcd}[row sep=large]
\displaystyle U(3) \times U(2) \arrow{d}{\displaystyle \langle \Phi \rangle} \\
\displaystyle \frac{SU(3)_c \times SU(2)_L \times U(1)_Y}{\mathbb{Z}_6} \arrow{d}{\displaystyle \langle H \rangle} \\
\displaystyle \frac{SU(3)_c \times U(1)_\text{em}}{\mathbb{Z}_3}
\end{tikzcd}
\end{displaymath}
Finally, let us remark that the minimal particle content reproduces exactly the SM Lagrangian, with the addition of a complex singlet fermion and scalar. The only unknown is whether the singlet scalar $\Phi$ interacts with the right-handed neutrinos $\nu_R$. This depends on the $\Phi$ charges, which are not totally fixed in our setup. \\

\noindent \textbf{Beyond the minimal setup.} Up to now, we have only considered the minimal consistent model from both the theoretical and phenomenological points of view. Now we will consider natural extensions of our setup that may address some open issues in the SM. \\

\textit{Neutrino masses--} Our model requires the presence of right-handed neutrinos, which couple to the SM lepton doublet through the Yukawa term $y_\nu \, \bar{L} \tilde{H} \nu_R$. As a result, neutrino masses are naturally generated. Moreover, the right-handed neutrinos are neutral under hypercharge and carry only a non-zero charge under the additional $U(1)$ symmetry. Consequently, both the nature of neutrino masses and the origin of their smallness depend on the way in which this extra $U(1)$ factor is broken. We identify three possible scenarios:
\begin{enumerate}
\item[(i)] If $\Phi$ carries charges that are twice those of $\nu_R$, a Yukawa term of the form
\begin{equation}
\Delta L_Y = \frac{\lambda}{2} \, \bar{\nu}_R^c \Phi^* \nu_R + \hc
\end{equation}
is allowed. Upon symmetry breaking, this term induces a Majorana mass for the right-handed neutrinos, $M_R = \lambda \, \langle \Phi \rangle$, reproducing the classical \textbf{type-I seesaw}~\cite{Minkowski:1977sc,Yanagida:1979as,Mohapatra:1979ia,GellMann:1979vob,Schechter:1980gr}.
\item [(ii)] If $\Phi$ carries the same charges as $\nu_R$, neutrinos are in principle Dirac fermions. However, this can be easily challenged. The model always allows for the inclusion of a purely singlet fermion, $\psi$. If this field is included, one can write the new Yukawa terms
\begin{align}
\Delta \L_Y = y_\psi \, \bar{\nu}_R^c \Phi \, \psi + \frac{\mu}{2} \, \bar{\psi}^c \psi +\hc \, ,
\end{align}
which lead to the \textbf{inverse seesaw}~\cite{Mohapatra:1986bd}.
\item [(iii)] Any other choice for the charges of $\Phi$ leads to Dirac neutrinos~\cite{Hirsch:2017col}.
\end{enumerate}

\textit{Dark matter--} Our model does not have a viable dark matter candidate, but it is straightforward to include one within a minimal extension \textit{without additional symmetries}. The spontaneous breaking of the new $U(1)$ symmetry may yield a residual $\Z_m$ discrete symmetry, with $m$ determined by the charge of $\Phi$, which can stabilize additional fields. This would provide a dynamical explanation for the origin of dark matter and constitute a very economical extension of our model. The generation of remnant discrete symmetries from the breaking of a gauge $U(1)$ factor was first explored in~\cite{Krauss:1988zc} and later exploited in the context of dark matter in several works, see for instance~\cite{Guo:2015lxa,AristizabalSierra:2015vqb}. \\

\textit{Unification--} Our fundamental idea of gauging all $U(1)$ redundancies by promoting the $SU(N)$ groups to $U(N)$ goes beyond the SM and can be extended to grand unified theories. For instance, applying this idea to the Georgi–Glashow $SU(5)$ model~\cite{Georgi:1974sy} leads to
\begin{equation}
U(5) \cong \frac{SU(5) \times U(1)}{\mathbb{Z}_5} \, ,
\end{equation}
which actually coincides with flipped $SU(5)$~\cite{Barr:1981qv}, provided the charge constraint derived from the $\mathbb{Z}_5$ quotient is respected. Moreover, $U(3) \times U(2)$ is a subgroup of $U(5)$. Therefore, it is not surprising that our $U(3) \times U(2)$ model can be naturally unified into a $U(5)$ model that can be identified with conventional flipped $SU(5)$. More generally, we note that our hypothesis implies that any grand unified theory based on an $SU(N)$ group should be extended to a $U(N)$ group, unifying all fundamental interactions with a single gauge group described by two distinct gauge couplings. \\

\noindent \textbf{Final discussion.} In this work we have taken the gauge principle to its logical limit: given an $SU(N)$ local transformation, the associated phase, usually a global symmetry, should become a local one by promoting $SU(N)$ to $U(N)$. We have shown how applying this idea to the SM framework one can explain some of its \textit{ad hoc} assumptions, such as charge quantization and the observed hypercharge values. Moreover, consistency of our setup requires the addition of right-handed neutrinos and leads to the existence of an additional $U(1)$ factor that can be identified with $B-L$, which in this way becomes a local gauge symmetry, instead of an unexplained accidental and global one, as in the SM.

In the particle physics literature, Lie groups are frequently invoked where only their Lie algebras are actually relevant, leading to a persistent and misleading confusion of the two. For instance, it is usually claimed that Glashow's seminal paper~\cite{Glashow:1961tr} identified the electroweak group, $SU(2)_L \times U(1)_Y$. However, this is not true: it identified the electroweak algebra, $\mathfrak{su}(2)_L \oplus \mathfrak{u}(1)_Y$. In fact, even today the gauge group of the SM is not fully determined~\cite{ORaifeartaigh:1986agb,Hucks:1990nw,Tong:2017oea}.

We point out that our $U(3) \times U(2)$ model cannot be distinguished experimentally from a model based on the local symmetry $SU(3)_c \times SU(2)_L \times U(1)_a \times U(1)_b$, provided the representations under the latter are consistent with their embedding in $U(3) \times U(2)$. This is simply because their respective algebras are isomorphic, so the resulting models are equivalent from a mathematical point of view. Therefore, the standard phenomenological avenues commonly explored in $U(1)$ extensions of the SM, such as $Z'$ searches at colliders and neutral gauge boson mixing, are also present in our framework. Their detailed implications, however, depend on the specific realization of the model (for instance, on the $U(1)$ charges of $\Phi$) and will be investigated in future work. That said, there are two crucial differences. First, our construction provides a symmetry-based explanation for the features of the resulting $U(1)$ extension of the SM, which would otherwise be completely ad hoc. And second, there are valid representations of $SU(3)_c \times SU(2)_L \times U(1)_a \times U(1)_b$ that cannot exist in $U(3) \times U(2)$. Therefore, even though our setup cannot be \textit{verified}, it can be \textit{falsified} by discovering one of these forbidden representations. This is how symmetries work: one can never confirm that we live in a symmetric universe, except by observing Nature's remarkable tendency to comply with our principles.

Our main message, to fully embrace the gauge principle, extends far beyond the specific $U(3) \times U(2)$ model presented in this paper. As noted, when applied to grand unification, it naturally leads to flipped $SU(5)$, which can be identified with $U(5)$. Our approach can be directly extended to a variety of other constructions. By adopting this perspective, we anticipate significant advancements in model building, with the potential to reveal novel avenues for New Physics yet to be uncovered.

\begin{center}
\vspace*{0.5cm}
\textbf{Acknowledgements}
\end{center}

The authors are grateful to Michal Malinsk\'y and the AHEP group for interesting discussions. Work supported by the Spanish grants PID2023-147306NB-I00, CNS2024-154524 and CEX2023-001292-S (MICIU/AEI/10.13039/501100011033), as well as CIPROM/2021/054 (Generalitat Valenciana). The work of AHB is supported by the grant No. CIACIF/2021/100 (also funded by Generalitat Valenciana). The work of JPS is supported by the grant CIACIF/2022/158 (Generalitat Valenciana).

%-------------------
% Bibliography
%-------------------
\bibliographystyle{apsrev4-2}
\bibliography{mybib.bib}

\clearpage

%-------------------
% End matter
%-------------------

\section*{End Matter}

In the End Matter we collect some basic results for $U(N)$ groups and discuss the cancellation of anomalies in the $U(3) \times U(2)$ model. \\

\noindent \textbf{Relation between $\boldsymbol{U(N)}$ and $\boldsymbol{SU(N) \times U(1)}$.} We first derive the isomorphism relation
\begin{equation}
  U(N) \cong \frac{SU(N) \times U(1)}{\Z_N}  
\end{equation}
and the restriction that it implies over the $U(1)$ charges. Let us start by proving the isomorphism. Consider the following group homomorphism $f : SU(N) \times U(1) \to U(N)$ such that $f(s \, , u ) = u \,  s$ with $u \in U(1)$ and $s \in SU(N)$. This $f$ is surjective. Thus, using the \textit{first isomorphism theorem} one finds 
\begin{equation}
    U(N) \cong \frac{SU(N) \times U(1)}{\text{ker} (f)} \; .
\end{equation}
The kernel of $f$ is simply given by
\begin{align}
    f(s\, , u) = u \, s = \id_N \, \Rightarrow \quad s = u^{-1} \, \id_N \; ,
\end{align}
with $u \mid u^N = 1$ since $\text{det}(s)=1$. Hence,
\begin{equation}
    \text{ker}(f) = \big\{ \left(z^{-1} \id_N \, , z \right) \mid z^N = 1 \big\} \cong \Z_N \, .
\end{equation}
Thus, finally, we recover 
\begin{equation}
    U(N) \cong \frac{SU(N) \times U(1)}{\mathbb{Z}_N} \; .
\end{equation}
Now we consider the irreducible representations of $U(N)$. Let \textbf{R}$_q$ be a representation of the group $G \equiv SU(N) \times U(1)$. \textbf{R}$_q$ is a representation of the quotient group iff  $\Z_N$ acts trivially under \textbf{R}$_q$. This condition reads
\begin{equation}
   \id_N =  R \left(z^{-1} \, \id_N \, , z\right) = (z^{-1} )^{m} \, z^q \, \id_N \, ,
\end{equation}
which implies
\begin{equation}
    q = m + k \, N \, \mid \, k \in \Z \, ,
\end{equation}
where $q$ and $m$ are the N-alities of the $U(1)$ and $SU(N)$ representations, respectively. In the particular cases of $U(2)$ and $U(3)$ we can express the relation as
\begin{align}
   q_2 = 2 \, j +2 \,n_2 \; , && q_3 = p-q +3 \,n_3 \; ,
\end{align}
where $n_2, n_3 \in \mathbb{Z} $ and $2 \, j$ (with $j$ the usual isospin of $SU(2)$ representations) and $(p, q)$ are the Dynkin labels for $SU(2)$ and $SU(3)$, respectively. \\

\noindent \textbf{Covariant derivative.} Let us consider a theory with $G$ as the gauge symmetry group. We define the covariant derivative as
\begin{equation}
    D_\mu = \partial_\mu - i \, g_a \,A_\mu^a \, T^a \, ,
\end{equation}
where we assign a different gauge coupling $g_a$ to each group generator $T^a$. Under $G$, matter and gauge fields transform with $U=e^{i \alpha_a T^a}$:
\begin{align}
    \phi &\to U\,  \phi \simeq \left(1 + i \alpha_a T^a \right)\,  \phi + \O\left(\alpha^2 \right), \\
    A_\mu^a T^a &\to U\, A_\mu^a T^a\, U^{-1} - \frac{i}{g}\, \left( \partial_\mu U \right) \,  U^{-1} \simeq \nonumber\\
    &\simeq A_\mu^a T^a + \frac{ \partial_\mu \alpha^a}{g}  T^a - f^{b c a} \, \alpha^b A_\mu^c \, T^a + \O(\alpha^2) .
\end{align}
It is straightforward to show that imposing the covariant condition 
\begin{equation}
D_\mu \psi \to U \, D_\mu \psi
\end{equation} 
leads to
\begin{equation}
    A_\mu^a \alpha_b \left(g_a - g_c \right) f^{abc} T^c \phi = 0 \; .
\end{equation}
Thus, the most general Lagrangian allows for different gauge couplings for those generators that commute with all the others, even if they belong to the same gauge group. This is, of course, equivalent to stating that the gauge physics is determined by the Lie algebra.\\

\noindent \textbf{Anomally cancellation.} In our model, perturbative anomalies arise from triangle diagrams involving $U(3)^3$, $U(2)^3$, $U(3)^2 \, U(2)$, $U(2)^2 \, U(3)$, $U(3) \, G^2$, and $U(2) \, G^2$. Since the generators of $U(N)$ consist of those of $SU(N)$ plus the identity, this is equivalent to imposing the cancellation of the $SU(3)^3$, $SU(3)^2 \, U(1)_3$, $U(1)_3^3$, $SU(2)^3$, $SU(2)^2 \, U(1)_2$, $U(1)_2^3$, $SU(3)^2 \, U(1)_2$, $SU(2)^2 \, U(1)_3$, $U(1)_3 \, G^2$, and $U(1)_2 \, G^2$ anomalies, since $\mathrm{Tr}\, T^a = 0$ for any $SU(N)$ generator $T^a$.

We assume that all fermions in the model transform either in the fundamental or in the singlet representation of $SU(N)$. Moreover, since $SU(2)$ is the SM $SU(2)_L$, we impose that right-handed fermions transform in the singlet representation of $SU(2)$.  As usual, the condition arising from $SU(3)^3$ requires an equal number of triplets and antitriplets, or equivalently, the same number of left- and right-handed triplets, whereas the $SU(2)^3$ anomaly vanishes identically since all representations of $SU(2)$ are real or pseudoreal. For the $SU(3)^2 \, U(1)_3$ anomaly, each triplet contributes a $U(1)_3$ charge of the form $1+3n$. The corresponding condition reads
\begin{align}
\sum_{i}^{N_{3L}} (1+3 n_i) - \sum_{i}^{N_{3R}} (1+3 m_i) = 0 \; .
\end{align}
Imposing that $n_i$ and $m_i$ are integers, we obtain
\begin{align}
m_{N_{3R}} &= \frac{N_{3L}-N_{3R}}{3} + \sum_{i}^{N_{3L}} n_i - \sum_{i}^{N_{3R}-1} m_i \nonumber \\
&\Rightarrow \; N_{3L} = 3 \, N_{3R} + k \; , \; k \in \mathbb{Z} \; .
\end{align}
Similarly, from the $SU(2)^2 \, U(1)_2$ anomaly we find
\begin{align}
\sum_{i}^{N_{2L}} (1+2 n_i) = 0 \quad \Rightarrow \quad N_{2L} = 2 \, k \; , \; k \in \mathbb{Z} \; .
\end{align}
On the other hand, the $U(1)_i \, G^2$ condition always includes $SU(3)^2 \, U(1)_i$ and $SU(2)^2 \, U(1)_i$ constraints, since all the $SU(N)$ representations contribute to the gravitational anomaly. This implies that the contributions from singlet and fundamental representations of $SU(N)$ must cancel independently. In the case of $U(1)_2 \, G^2$, this requires the presence of complete singlet fermions under both $SU(N)$. In particular, since the gauge group $U(3)\times U(2)$ contains two $U(1)$ factors, two complete singlets are required. Altogether, we find that the minimal fermion content coincides with that of a single generation of the Standard Model. For this minimal content, the conditions arising from the $U(N)^3$ anomalies read
\begin{equation}
\begin{aligned}
SU(3)^2 \, U(1)_i \; &: \; 2 q_i - d_i - u_i = 0 \; ,\\
SU(2)^2 \, U(1)_i \; &: \; 3 q_i - l_i = 0 \; ,  \\
G^2 \, U(1)_i \; &: \; 6 q_i + 2 l_i - 3 d_i - 3 u_i - e_i - \nu_i = 0 \; ,\\
U(1)_i^3 \; &: \; 6 q_i^3 + 2 l_i^3 - 3 d_i^3 - 3 u_i^3 - e_i^3 - \nu_i^3 = 0  \; .
\end{aligned}
\end{equation}
These equations are solved by
\begin{equation}
l_i = -3 q_i \; , \;\; d_i = 2 q_i - u_i \; , \;\; e_i = -4 q_i + u_i \; , \;\; \nu_i = -2 q_i - u_i \; .
\end{equation}
Finally, it is straightforward to verify that this solution is also compatible with the mixed anomalies $U(3)^2 \, U(2)$ and $U(2)^2 \, U(3)$.

\end{document}